\documentclass{aa}
\usepackage{graphicx}
\usepackage{txfonts}
\usepackage[usenames,dvipsnames]{color}
\usepackage{natbib,twoopt}
\usepackage[breaklinks=true,linktocpage=true,backref=true,pagebackref=false,bookmarks=true,bookmarksnumbered=False,colorlinks=true,linkcolor=NavyBlue,citecolor=NavyBlue,urlcolor=NavyBlue]{hyperref} %% to avoid \citeads line fills
\bibpunct{(}{)}{;}{a}{}{,}             %% natbib format for A&A and ApJ
\makeatletter
  \newcommandtwoopt{\citeads}[3][][]{\href{http://adsabs.harvard.edu/abs/#3}%
    {\def\hyper@linkstart##1##2{}%
     \let\hyper@linkend\@empty\citealp[#1][#2]{#3}}}
  \newcommandtwoopt{\citepads}[3][][]{\href{http://adsabs.harvard.edu/abs/#3}%
    {\def\hyper@linkstart##1##2{}%
     \let\hyper@linkend\@empty\citep[#1][#2]{#3}}}
  \newcommandtwoopt{\citetads}[3][][]{\href{http://adsabs.harvard.edu/abs/#3}%
    {\def\hyper@linkstart##1##2{}%
     \let\hyper@linkend\@empty\citet[#1][#2]{#3}}}
  \newcommandtwoopt{\citeyearads}[3][][]%
    {\href{http://adsabs.harvard.edu/abs/#3}i
    {\def\hyper@linkstart##1##2{}%
     \let\hyper@linkend\@empty\citeyear[#1][#2]{#3}}}
\makeatother

\DeclareSymbolFont{UPM}{U}{eur}{m}{n}
\SetSymbolFont{UPM}{bold}{U}{eur}{b}{n}
\DeclareMathSymbol{\umu}{0}{UPM}{"16}
\def\micron{\hbox{$\umu$m}}

\begin{document} 
   \title{Embedded AGN and star formation in the central 80 pc of IC 3639\thanks{Based on observations collected at the European Southern Observatory, Chile, programmes 070.B-0393, 088.D--0005 and 088.B--0809.}}

   %\subtitle{}
   \author{J.\,A. Fern\'andez-Ontiveros \inst{1,2,3}
          \and
          K.\,R.\,W. Tristram \inst{4}
          \and
          S. H\"onig \inst{5}
          \and
          P. Gandhi \inst{5}
          \and
          G. Weigelt \inst{3}
          }

   \institute{Instituto de Astrof\'isica de Canarias (IAC), C/V\'ia L\'actea s/n, E--38205 La Laguna, Tenerife, Spain\\
              \email{j.a.fernandez.ontiveros@gmail.com, jafo@iac.es}
              \and
              Universidad de La Laguna (ULL), Dpto. Astrof\'isica, Avd. Astrof\'isico Fco. S\'anchez s/n, E--38206 La Laguna, Tenerife, Spain
              \and
              Max-Planck-Institut f\"ur Radioastronomie (MPIfR), Auf dem H\"ugel 69, D--53121 Bonn, Germany
              \and
              European Southern Observatory, Alonso de C\'ordova 3107, Vitacura, Santiago, Chile
              \and
              School of Physics \& Astronomy, University of Southampton, Southampton SO17 1BJ, UK
   }

   %   \date{Received September 15, 1996; accepted March 16, 1997}
   \date{\today}

% \abstract{}{}{}{}{} 
% 5 {} token are mandatory

   \abstract
  {}
  {Our goal is to probe the inner structure and the nature of the mid-IR emission in the active galaxy IC\,3639, which hosts a Seyfert 2 nucleus and shows signatures of strong star-forming activity.}
  {We use interferometric observations in the \textit{N}-band with VLTI/MIDI to resolve the mid-IR emission of this nucleus. The origin of the nuclear infrared emission is determined from: 1) the comparison of the correlated fluxes from VLTI/MIDI with the fluxes measured at subarcsec resolution (VLT/VISIR, VLT/ISAAC); 2) diagnostics based on IR fine-structure line ratios, the IR continuum emission, IR bands produced by polycyclic aromatic hydrocarbons (PAH) and silicates; and 3) the high-angular resolution spectral energy distribution.}
  {A large fraction of the total mid-IR emission of IC\,3639 is produced in the innermost $\lesssim 80\, \rm{pc}$ with only $\sim$\,$1\%$ of the total luminosity released in the UV/optical range. The unresolved flux of IC\,3639 is $90 \pm 20\, \rm{mJy}$ at $10.5\, \rm{\micron}$, measured with three different baselines in VLTI (UT1--UT2, UT3--UT4, and UT2--UT3; $46$--$58\, \rm{m}$), making this the faintest measurement so far achieved with mid-IR interferometry. The correlated flux is a factor of $3$--$4$ times fainter than the VLT/VISIR total flux measurement. The observations suggest that most of the mid-IR emission has its origin on spatial scales between $10$ and $80\, \rm{pc}$ ($40$--$340\, \rm{mas}$). The emission confined within the inner $80\, \rm{pc}$ is either dominated by a starburst component or by the AGN core. The brightness distribution could be reproduced by a single component associated to the AGN, although this scenario would imply a very extended dust distribution when compared to other nearby Seyfert galaxies detected with MIDI. The extended component could also be associated to polar dust emission, i.e. a dusty wind blown by the AGN. However, a mixed contribution dominated by the star formation component over the AGN is favoured by the diagnostics based on ratios of IR fine-structure emission lines, the shape of the IR continuum, and the PAH and silicate bands.}
  {A composite AGN-starburst scenario is able to explain both the mid-IR brightness distribution and the IR spectral properties observed in the nucleus of IC\,3639. The nuclear starburst would dominate the mid-IR emission and the ionisation of low-excitation lines (e.g. [\ion{Ne}{ii}]$_{12.8 \rm{\mu m}}$) with a net contribution of $\sim$\,$70\%$. The AGN accounts for the remaining $\sim$\,$30\%$ of the mid-IR flux, ascribed to the unresolved component in the MIDI observations, and the ionisation of high-excitation lines (e.g. [\ion{Ne}{v}]$_{14.3 \rm{\mu m}}$ and [\ion{O}{iv}]$_{25.9 \rm{\mu m}}$).
    % If this scenario is confirmed, the case of IC\,3639 would be in contrast with the quenching of star-formation activity by AGN feedback, and suggests that AGN activity and star formation can coexist in a dust-embedded phase within the inner $80\, \rm{pc}$ of this galaxy.
  }


   \maketitle

%
%~~~~~~~~~~~~~~~~~~~~~~~~~~~~~~~~~~~~~~~~~~

\section{Introduction}\label{intro}

The central regions of nearby galaxies are excellent laboratories to probe the connection between Active Galactic Nuclei (AGN) and starburst activity. The $M$--$\sigma$ relation suggests a coupling between both phenomena, since the growth of the black hole (BH) correlates with the properties of the stellar bulge \citepads{1995ARA&A..33..581K,2000ApJ...539L..13G}. This coexistence has been identified up to $z \sim 2$, where strong AGN activity co-evolve with star formation in galaxies, both embedded in a dust enshrouded phase mostly obscured to optical wavelengths, but revealed in the infrared (IR) range \citepads{2014ARA&A..52..415M}. The main uncertainties arise when we try to address the interplay and evolution of AGN and star formation, for instance if gas heating and outflows from the AGN quench the star-formation process and further AGN activity, self-regulating the growth of the BH \citepads{2003MNRAS.339..289S,2005Natur.433..604D,2007ApJ...665.1038C}. Alternatively, there might be a triggering of the AGN activity produced by the starburst itself, driving the material into the innermost parsecs to fuel the supermassive BH \citepads{2008ApJ...681...73K,2013A&A...560A..34W}.

\begin{table*}
\caption{Optical, IR photometry, and IR emission lines and spectral features measured for the nucleus of IC\,3639. Optical photometric fluxes have been corrected by Galactic extinction ($E(B - V) = 0.06\, \rm{mag}$; \citeads{2011ApJ...737..103S,1989ApJ...345..245C}). In the subsequent analysis, data were also corrected for redshift.}\label{flux}
\centering
\begin{tabular}{lcccc}
Instrument/Filter & $\lambda$ ($\rm{\micron}$) & $\Delta\lambda$ ($\rm{\micron}$) & Flux (mJy) & Aperture \\
\hline \\[-0.3cm]
  FOC/F210M & $0.22$ & $0.04$ & $(17 \pm 2) \times 10^{-3}$ & $0\farcs14$ \\ %$(175 \pm 8) \times 10^{-4}$
  WFPC2/F606W & $0.60$ & $0.15$ & $0.22 \pm 0.02$ & $0\farcs14$ \\ %$(22 \pm 2) \times 10^{-2}$
  ISAAC/\textit{L}-band & $3.78$ & $0.58$ & $18.9 \pm 0.2$ & $0\farcs50$ \\ %$18.7 \pm 0.2$
  ISAAC/\textit{M}-band & $4.66$ & $0.10$ & $32 \pm 1$ & $0\farcs50$ \\
  T-ReCS/\textit{N}-band & $10.36$ & $2.64$ & $305 \pm 9$\tablefootmark{a} & $0\farcs40$ \\ % $304 \pm 9$
  VISIR/PAH2 & $11.25$ & $0.59$ & $326 \pm 17$\tablefootmark{b} & $0\farcs40$ \\
  VISIR/NEII & $12.81$ & $0.10$  & $532 \pm 30$ & $0\farcs50$ \\
  VISIR/NEII\_2 & $13.04$ & $0.22$ & $543 \pm 29$\tablefootmark{b} & $0\farcs40$ \\
  \hline \\[-0.3cm]
  MIDI/\textit{N}-band (UT2--UT3) & $10.5$ & $2.5$  &  $84 \pm 17$\tablefootmark{c} & $0\farcs52$ \\
  MIDI/\textit{N}-band (UT1--UT2) & $10.5$ & $2.5$  &  $95 \pm 21$\tablefootmark{c} & $0\farcs52$ \\
  MIDI/\textit{N}-band (UT3--UT4) & $10.5$ & $2.5$  &  $90 \pm 17$\tablefootmark{c} & $0\farcs52$ \\
  \hline \\[-0.1cm]
  Instrument              && Line & Flux ($\times 10^{-17}\, \rm{W\,m^{-2}}$)         & Slit width \\
  \hline \\[-0.3cm]
  ESO-MPG2.2m/EFOSC2      && [\ion{O}{iii}]$^{\rm obs}_{5007\, \rm{\AA}}$ & $20.3 \pm 0.2$\tablefootmark{d} & $\sim 1''$ \\
  ESO-MPG2.2m/EFOSC2      && [\ion{O}{iii}]$^{\rm cor}_{5007\, \rm{\AA}}$ & $185.7 \pm 1.8$\tablefootmark{d} & $\sim 1''$ \\
  \textit{Spitzer}/IRS    && PAH$_{11.3\, \rm{\mu m}}$                   & $111$\tablefootmark{e} & $4\farcs7 \times 11\farcs3$ \\
  \textit{Spitzer}/IRS    && [\ion{Ne}{ii}]$_{12.8\, \rm{\mu m}}$        & $45.2 \pm 0.5$\tablefootmark{e} & $4\farcs7 \times 11\farcs3$ \\
  \textit{Spitzer}/IRS    && [\ion{Ne}{v}]$_{14.3\, \rm{\mu m}}$         & $11.2 \pm 0.6$\tablefootmark{e} & $4\farcs7 \times 11\farcs3$ \\
  \textit{Spitzer}/IRS    && [\ion{Ne}{iii}]$_{15.6\, \rm{\mu m}}$       & $27.0 \pm 0.4$\tablefootmark{e} & $4\farcs7 \times 11\farcs3$ \\
  \textit{Spitzer}/IRS    && [\ion{O}{iv}]$_{25.9\, \rm{\mu m}}$         & $21.2 \pm 0.6$\tablefootmark{e} & $11\farcs1 \times 22\farcs3$ \\
  \textit{Herschel}/PACS  && [\ion{O}{iii}]$_{88\, \rm{\mu m}}$          & $41.9 \pm 1.7$\tablefootmark{f} & $28\farcs2 \times 28\farcs2$ \\
  VISIR                   && [\ion{S}{iv}]$_{10.5\, \rm{\mu m}}$         & $7.1 \pm 1.6$  & $0\farcs52$ \\
  VISIR                   && PAH$_{11.3\, \rm{\mu m}}$                   & $41.4 \pm 0.4$ & $0\farcs52$ \\
  VISIR                   && [\ion{Ne}{ii}]$_{12.8\, \rm{\mu m}}$        & $33.4 \pm 1.0$ & $0\farcs52$ \\[0.1cm]
  \hline
\end{tabular}
\tablefoot{
\tablefoottext{a}{PSF photometry from \citetads{2014MNRAS.439.1648A}.}
\tablefoottext{b}{Photometry from \citetads{2009A&A...502..457G}.}
\tablefoottext{c}{Correlated flux in the UT2--UT3 (46\,m), UT1--UT2 (56.5\,m), and UT3--UT4 (57.8\,m) baselines.}
\tablefoottext{d}{Line fluxes from \citetads{2003ApJS..148..353T}, including both observed and corrected fluxes from the measured Balmer decrement of $6.1$, assuming an intrinsic decrement of $3.0$ \citepads{1999ApJS..121..473B}.}
\tablefoottext{e}{Line fluxes from \citetads{2008ApJ...676..836T}.}
\tablefoottext{f}{Line flux from \citetads{2016ApJS..226...19F}.}
}
\end{table*}

Although AGN and starburst activity show coexistence in time, with short delays between them ($\sim$$100$--$200\, \rm{Myr}$, \citeads{2007ApJ...671.1388D,2010MNRAS.405..933W}), they are typically identified at very different spatial scales. While the AGN releases its energy within a region of a few Schwarzschild radii, nuclear star-forming regions have been spatially resolved from $\sim 100\, \rm{pc}$ to kiloparsec-size rings \citepads{1998ApJ...505..174G,2000AJ....120.1325W,2005A&A...429..141K}. Traces of star-formation have also been detected in the inner few tens of parsecs of AGN, possibly associated with the torus \citepads[e.g.][]{2007ApJ...671.1388D,2014ApJ...780...86E}. Such a mechanism is sometimes invoked in torus models in order to support its vertical structure, for example, through supernovae (SNe) explosions (\citeads{2002ApJ...566L..21W}, \citeads{2008A&A...491..441V}, \citeads{2009MNRAS.393..759S}). On the other hand, interferometric observations in the IR of nearby AGN have changed our picture of the dusty media surrounding the central engine (the torus; e.g. \citeads{2004A&A...425...77W}, \citeads{2004Natur.429...47J}, \citeads{2013A&A...558A.149B}, \citeads{2013ApJ...771...87H}, \mbox{\citeads{2014A&A...563A..82T}}, \citeads{2014A&A...565A..71L}), and also the cold molecular gas revealed by ALMA (\citeads{2016ApJ...823L..12G}, \citeads{2016ApJ...829L...7G}, \citeads{2017arXiv170205458A}). A theoretical scenario to describe the observed characteristics has been proposed recently by \citetads{2017ApJ...838L..20H}, including an extended dusty wind dominating the mid-IR flux and a compact geometrically-thin disc component dominating the near-IR bump. However, it is not clear what is the effect of star formation happening in the innermost few parsecs of a galaxy, which mechanisms can regulate the interplay with the central BH, and what is the impact on the obscuration and the dust distribution in the torus.

\begin{figure*}
  \sidecaption
  \includegraphics[width = 12cm]{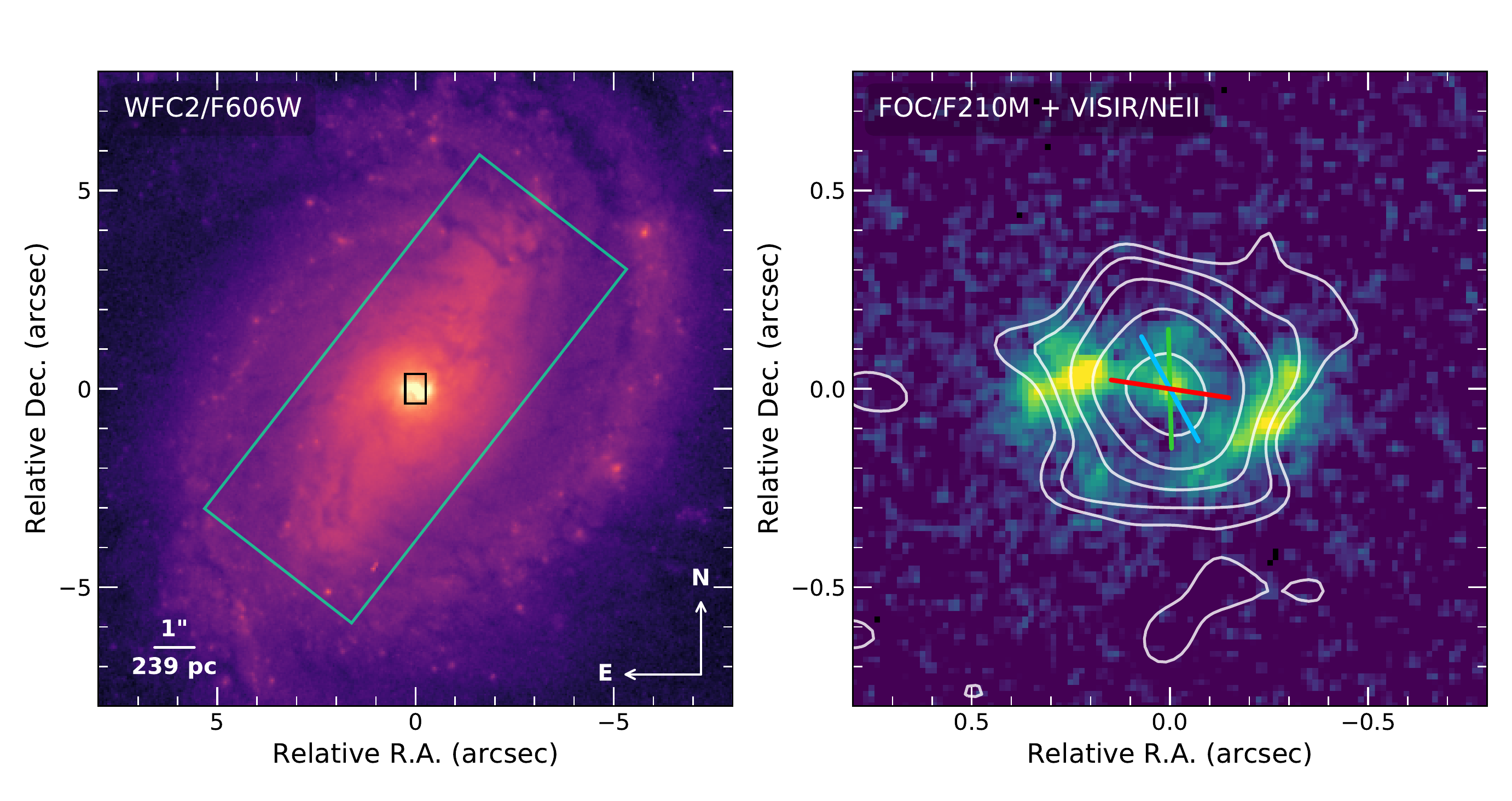}
  \caption{\textit{Left:} \textit{HST}/WFC2 image of IC\,3639 acquired with the filter F606W. The black rectangle represents the size of the slit used to extract the VISIR spectrum ($0\farcs75 \times 0\farcs52$), while the green square corresponds to the slit size and orientation of the \textit{Spitzer}/IRS-SH observations ($9.9$--$19.5\, \rm{\micron}$, $4\farcs7 \times 11\farcs3$). \textit{Right:} \textit{HST}/FOC image acquired with the filter F210M from \citetads{1998ApJ...505..174G}. The white contours correspond to the VISIR/SPEC NEII filter image. The orientation of the different baselines used for MIDI observations are indicated by the blue (UT2--UT3), green (UT1--UT2), and red (UT3--UT4) lines.\vspace{0.9cm}}\label{mirmap}
\end{figure*}

IC\,3639 is a source well suited to explore the AGN-starburst connection. This Seyfert 2 nucleus shows mixed characteristics with star formation activity \citepads{1998ApJ...505..174G}, a relatively bright mid-IR core ($\sim$\,$320\, \rm{mJy}$; \citeads{2014MNRAS.439.1648A}), and a broad H$\alpha$ component in polarised light suggesting the presence of a hidden type 1 nucleus \citepads{1997Natur.385..700H,2001MNRAS.327..459L}. Its nucleus is strongly absorbed in X-rays ($N_H > 3.6 \times 10^{24}\, \rm{cm^{-2}}$, i.e.\ Compton-thick; \citeads{2016ApJ...833..245B}). In the UV, images from the Faint Object Camera (FOC) on board the \textit{Hubble Space Telescope} (\textit{HST}) reveal an extended morphology along the east-west axis (\citeads{1998ApJ...505..174G}; see Fig.\,\ref{mirmap}, right panel). Three main regions are resolved by \textit{HST}/FOC: a fainter compact one --\,assumed to be the nucleus\,-- located in the centre, which is surrounded by two brighter and extended lobes at a distance of $\sim 0\farcs2$ ($48\, \rm{pc}$; $D = 49.4\, \rm{Mpc}$ from NED\footnote{$H_0 =  73\, \rm{km\,s^{-1}\,Mpc^{-1}}$, $\Omega_m = 0.27$, $\Omega_\Lambda = 0.73$.}). Still, it is not clear whether these lobes correspond to circumnuclear star-formation \mbox{\citepads{1998ApJ...505..174G}} or to the narrow-line region (NLR; \citeads{2007ApJ...656..105G}). Furthermore, a kinematic study in the UV performed by \citetads{2013ApJS..209....1F} to determine the NLR orientation was not conclusive for the IC\,3639. Finally, an extended radio jet has not been reported in the literature for this object, although extended radio emission at subarcsec scales was found by \citetads{2000MNRAS.314..573T}.

%~~~~~~~~~~~~~~~~~~~~~~~~~~~~~~~~~~~~~~~~~~
%~~~~~~~~~~~~~~~~~~~~~~~~~~~~~~~~~~~~~~~~~~

Due to the compactness of the emission in the mid-IR, the nucleus of IC\,3639 is an obvious target for the MID-infrared Interferometric instrument (MIDI) at the Very Large Telecope Interferometer (VLTI). In the \textit{N}-band, this source is unresolved at $0\farcs4$ resolution in VLT/VISIR\footnote{V\textsc{lt} Imager and Spectrometer for mid-InfraRed \citepads{2004Msngr.117...12L}.} images (see Fig.\,\ref{mirmap} right panel UV--IR comparison; also \citeads{2014MNRAS.439.1648A}). The point-like emission is ascribed to the UV and optical central peak, but no counterpart has been found in the mid-IR down to $\sim 15\, \rm{mJy}$ for the two bright UV peaks surrounding the nucleus \citepads{2009A&A...502..457G,2014MNRAS.439.1648A}. With a mid-IR flux of $\sim$\,$320\, \rm{mJy}$, IC\,3639 was one of the targets included in the MIDI AGN Large program\footnote{ESO program number 184.B-0832.}, a campaign performed with VLTI/MIDI. However, no interferometric signal could be detected for the nucleus of IC\,3639. A flux limit of $F_\nu < 150\, \rm{mJy}$ was derived for this target for a projected baseline length of $56\,\rm{m}$ using the UT1--UT2 configuration \mbox{\citepads{2013A&A...558A.149B}}. This motivated our targeted observation of this source using the experimental VLTI/MIDI ``no-track'' mode, which allows the detection of faint sources where fringes cannot be tracked.

The observations and the data reduction are explained in Section\,\ref{data}, the main results are presented in Section\,\ref{results}, and its main implications for the AGN-starburst connection are discussed in Section\,\ref{discussion}. Finally, in Section\,\ref{summary} we summarise the main conclusions from this study.

\section{Dataset}\label{data}
The new observations in the mid-IR were carried out with VLTI/MIDI and VLT/VISIR. These were complemented with archival mid-IR spectra from the Infrared Spectrograph on board \textit{Spitzer}, \textit{L} and \textit{M}-band images from VLT/ISAAC and \textit{HST} optical and UV images. Flux measurements of the AGN derived from these data, corrected for Galactic reddening, are included in Table\,\ref{flux}.

\subsection{Mid-IR observations}
IC\,3639 was observed with MIDI \citepads{2003Ap&SS.286...73L}, a Michelson dual-beam combiner for spectro-interferometric observations in the $8$--$13\, \rm{\micron}$ wavelength range, located at the Very Large Telescope Interferometer (VLTI), Cerro Paranal, Chile. Very faint fringes on IC\,3639 were detected under very stable atmospheric conditions on the nights of February 5th (UT1--UT2 configuration), and February 10th 2012 (UT3--UT4 and UT2--UT3). The projected baselines are $56.5\, \rm{m}$ (\textit{PA}\,$= 1.5^\circ$), $57.8\, \rm{m}$ ($81.4^\circ$), and $46.0\, \rm{m}$ ($28.6^\circ$), respectively. The faintness of this target in the \textit{N}-band prevents the use of the standard fringe tracking mode. Thus, the observations were performed using the no-tracking mode, i.e. the fringes were scanned blindly --\,with the group-delay tracking switched off\,-- close to the zero optical path delay, derived for the calibrator (HD\,108759). During the first night (UT1--UT2), a brighter AGN --\,IC\,4329A\,-- was observed close in time with the same instrumental configuration. The spectral properties of IC\,4329A are similar to those of IC\,3639 and hence allow us to robustly constrain the correlated flux of IC\,3639. For the second night (UT3--UT4 and UT2--UT3), the correlated flux was determined with the help of another bright target, V838\,Mon\footnote{This source was originally observed for a different science case under programme 088.D--0005.}.

The data of IC\,3639 is too faint to be processed by the standard MIDI data reduction software \textsc{mia+ews} \citepads{2004SPIE.5491..715J}. Instead, we developed further the method presented in \citetads{2012SPIE.8445E..1GB} in order to estimate the correlated flux of a faint source. To obtain an estimate of the broadband correlated flux from the resulting data, the signal of IC\,3639 was compared to that of the brighter sources, IC\,4329A and V838\,Mon. The latter were reduced by a certain factor and mixed with the noise recorded in IC\,3639 to mimic its signal. We found that the correlated flux of IC\,4329A ($F_\nu \sim 700\, \rm{mJy}$) needs to be reduced by a factor of $\sim 0.13 \pm 0.03$ to match the signal observed for IC\,3639, implying a correlated flux of $95 \pm 21\, \rm{mJy}$ for IC\,3639 on the UT1--UT2 baseline. V838\,Mon ($F_\nu \sim 800\, \rm{mJy}$) required a reduction by a factor of $\sim 0.11$ (UT3--UT4) and $0.10$ (UT2--UT3), corresponding to correlated fluxes of $90 \pm 17\, \rm{mJy}$ and $84 \pm 17\, \rm{mJy}$, respectively. With this method only an estimate of the broadband correlated flux can be obtained, the spectral information cannot be recovered.

\begin{figure*}
  \sidecaption
  \includegraphics[width = 12cm]{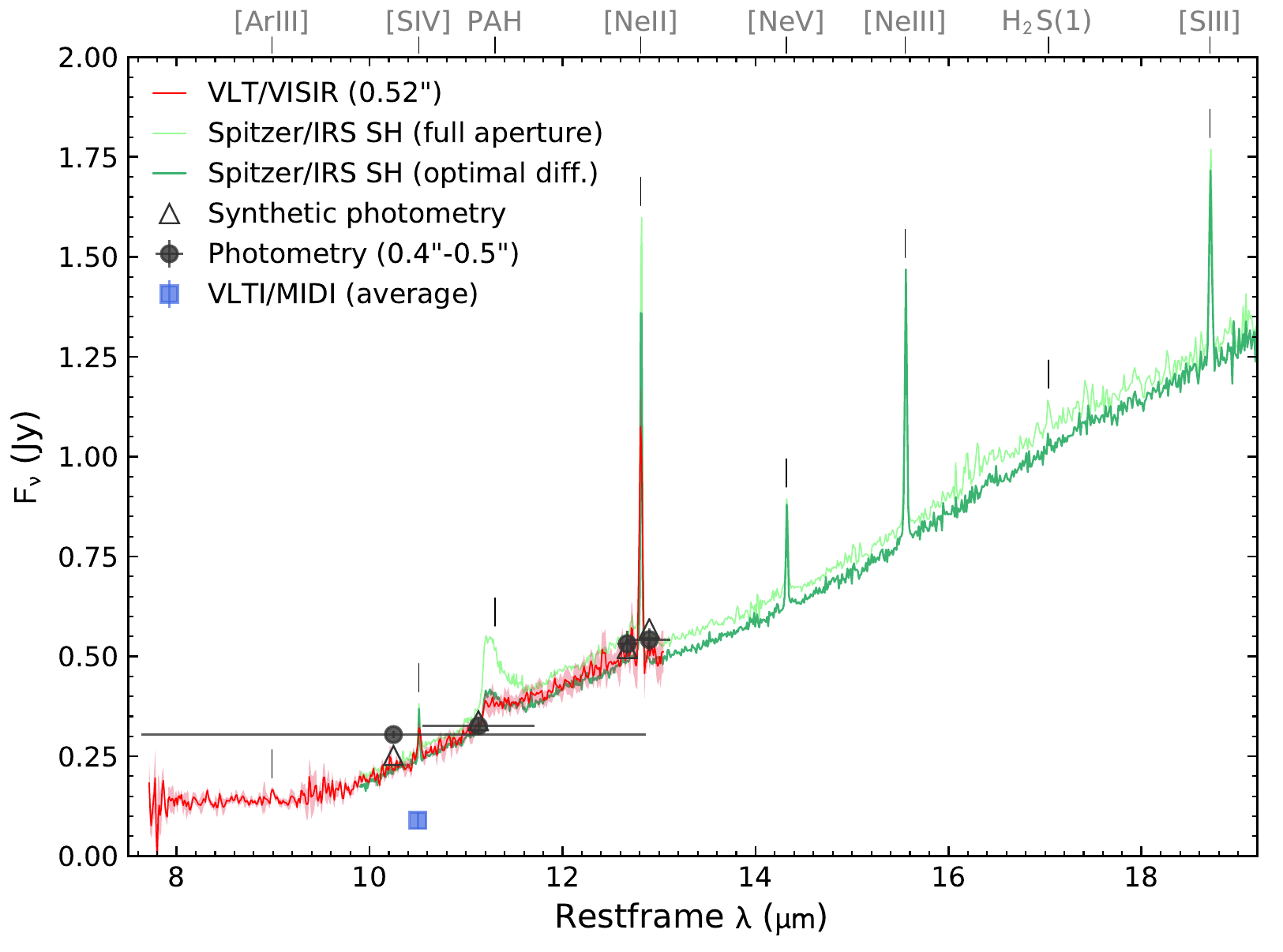}
  \caption{Mid-IR rest-frame spectrum of IC\,3639. In red, the VLT/VISIR spectrum acquired with a slit size of $0\farcs75 \times 0\farcs52$ and a spectral resolution of $R \sim 350$. In light green, the \textit{Spitzer}/IRS spectrum at $R \sim 600$ obtained from the full-aperture extraction (slit size of $4\farcs7 \times 11\farcs3$). In dark green, the optimal-difference extracted spectrum of the same observation (see text). Both IRS spectra were taken from the CASSIS atlas \citepads{2015ApJS..218...21L}. Black circles correspond to VLT/VISIR photometric measurements at $\sim 0\farcs5$ resolution ($96\, \rm{pc}$). The average correlated flux estimated from VLTI/MIDI ($\sim 90\, \rm{mJy}$), is represented by the blue square. Black open triangles correspond to synthetic photometry performed on the VISIR spectrum. Note that no scaling has been applied to the \textit{Spitzer}/IRS-SH spectrum, thus the match between the continuum level in both spectra suggests that the point-like source seen by VISIR dominates also the \textit{Spitzer}/IRS-SH spectrum. Emission lines stand out in the IRS spectrum due to the higher spectral resolution.\vspace{0.6cm}}\label{mirspec}
\end{figure*}

In parallel, low-resolution ($R \sim 350$) mid-IR spectra were acquired with VLT/VISIR on February 5th, 2012 ($8.5$ and $9.8\, \rm{\micron}$ spectral settings; calibrator HD\,102964) and March 8th 2012 ($11.4$ and $12.4\, \rm{\micron}$ spectral settings; calibrator HD\,168592). The data were reduced and calibrated using the ESO Common Pipeline Library (\textsc{CPL} v5.3.1, VISIR recipe v3.4.4), and post-processed using the methods described in \citetads{2010A&A...515A..23H}. The final spectrum was extracted from a $0\farcs75 \times 0\farcs52$ slit, which mimics the slit characteristics of MIDI (slit represented as a black rectangle in Fig.\,\ref{mirmap}, left panel; the spectrum is shown in red colour in Fig.\,\ref{mirspec}). Additionally, aperture photometry ($0\farcs5$ radius, background estimated in a $0\farcs6$--$0\farcs8$ annulus) of the science target was performed on the NEII filter acquisition image, taken simultaneously with the spectra ($\lambda = 12.81\, \rm{\micron}$, $\Delta\lambda = 0.21\, \rm{\micron}$; calibrator HD\,168592).

The continuum fluxes measured in the VISIR images are listed in Table\,\ref{flux}. The latter also includes the integrated fluxes measured in the VISIR spectrum for the [\ion{S}{iv}]$_{10.5\, \rm{\mu m}}$ and [\ion{Ne}{ii}]$_{12.8\, \rm{\mu m}}$ emission lines, and the PAH$_{11.3\, \rm{\mu m}}$ feature. In all cases, the continuum contribution has been subtracted using a linear interpolation to the continuum adjacent to the corresponding emission line or feature.

%~~~~~~~~~~~~~~~~~~~~~~~~~~~~~~~~~~~~~~~~~~
%~~~~~~~~~~~~~~~~~~~~~~~~~~~~~~~~~~~~~~~~~~

\subsection{Archival data}
In our analysis we included a mid-IR spectrum acquired with the InfraRed Spectrograph (IRS) on board the \textit{Spitzer Space Telescope} (PI: G. Fazio; AOR key: 17961984; see Fig.\,\ref{mirspec}), using the high-spectral resolution mode ($R \sim 600$). The latter covers the $9.9$--$37.1\, \rm{\micron}$ range with slit sizes of $4\farcs7 \times 11\farcs3$ (SH mode: $9.9$--$19.5\, \rm{\micron}$; green slit in Fig.\,\ref{mirmap}) and $11\farcs1 \times 22\farcs3$ (LH mode: $18.8$--$37.1\, \rm{\micron}$). Two different extractions of the same spectrum were taken from the Cornell Atlas of \textit{Spitzer}/Infrared Spectrograph Sources (CASSIS; \citeads{2015ApJS..218...21L}): the full-aperture extraction (light green spectrum in Fig.\,\ref{mirspec}), and the optimal-differential extraction (dark green spectrum in Fig.\,\ref{mirspec}). The former is obtained from the collapse of the 2D spectrum along the wavelength axis within the full-aperture slit ($4\farcs7 \times 11\farcs3$ for the SH range). In the optimal-differential method, the pipeline used in CASSIS takes advantage of the PSF profile and the two nod images to improve the S/N ratio and produce a background-subtracted spectrum (see \citeads{2015ApJS..218...21L}). This method is particularly suited for unresolved sources mixed with extended emission. Note that no offset scaling has been applied to the \textit{Spitzer}/IRS spectra, whose continuum level matches perfectly the VISIR spectrum in the $10$--$13\, \rm{\micron}$ range. Line fluxes measured by \citetads{2008ApJ...676..836T} in a full-aperture extraction of the \textit{Spitzer}/IRS high-spectral resolution (HR) spectrum are listed in Table\,\ref{flux}.

VLT/ISAAC \textit{L} and \textit{M}-band observations were taken from the ESO scientific archive (July 15th 2003, programme 070.B-0393) and are included in Table\,\ref{flux}. The data reduction was performed using the \textit{jitter} task within \textsc{eclipse} \citepads{1999ASPC..172..333D}, including background subtraction, registration, and combination of individual frames. Flux calibration was based in the near-IR standard HD\,106965 (UKIRT standards; \citeads{2003MNRAS.345..144L}). The photometry in the \textit{L} and \textit{M}-band images was extracted using a circular aperture of $0\farcs5$ radius (background estimated in a $0\farcs6$--$0\farcs8$ annulus).

Finally, reduced and calibrated images were retrieved from the \textit{HST} scientific archive, acquired with the filters F210M (FOC; $\lambda = 2179.4$\,\AA, $\Delta\lambda = 381.5$\,\AA) and F606W (WFPC2; $\lambda = 6001.3$\,\AA, $\Delta\lambda = 1498.8$\,\AA). This corresponds to the same \textit{HST} dataset as presented in \citetads{1998ApJ...505..174G}. The photometry in the \textit{HST} images was extracted using a circular aperture of $0\farcs14$ radius, subtracting the background estimated in a $0\farcs2$--$0\farcs3$ annulus (see Table\,\ref{flux}).

%~~~~~~~~~~~~~~~~~~~~~~~~~~~~~~~~~~~~~~~~~~

\section{Results}\label{results}

\subsection{Morphology}
As probed by the \textit{HST}/FOC image (\citeads{1998ApJ...505..174G}), the nucleus of IC\,3639 is an extended UV source, with a size of $0\farcs8 \times 0\farcs4$ ($192 \times 96\, \rm{pc^2}$). The nuclear emission is resolved into a substructure formed by three distinct regions: a central core likely associated with the AGN and two side lobes associated with star formation (see Fig.\,\ref{mirmap}). In the \textit{L}, \textit{M}, and \textit{N}-bands, the morphology shows only a point-like source at an angular resolution of $< 0\farcs4$ \citepads[see fig.\,3 in][]{2009A&A...502..457G}. This means that the UV side lobes do not show any counterpart in the mid-IR images (see Fig.\,\ref{mirmap}). Although \citetads{2014MNRAS.439.1648A} classify IC\,3639 as a marginally resolved source in the mid-IR NEII\_2 narrow-band filter image (\textsc{fwhm}$\sim 0\farcs39 \sim 91\, \rm{pc}$, \textit{PA} $\sim 79^\circ$), such extension is not detected in any of the other filters. Consequently, \citetads{2016ApJ...822..109A} classify the source as non-extended (\textsc{fwhm}$\lesssim 0\farcs34 \sim 80\, \rm{pc}$). Hereafter we will adopt $80\, \rm{pc}$ as a conservative upper limit to the size of the source, although the point-like morphology of this nucleus in VISIR images suggests a narrower flux distribution with \textsc{fwhm}$\lesssim 50\, \rm{pc}$ as a more plausible scenario. The two UV lobes are brighter than the nucleus in the F210M filter, but they become fainter at longer wavelengths in the F606W filter. Thus, the bright point-like mid-IR source is assumed to be the counterpart of the central UV/optical peak. The origin of the central optical peak is not clear, since IC\,3639 is a type 2 Compton-thick nucleus (\citeads{2005A&A...444..119G}) and thus the AGN should be covered by the obscuring torus in this range. Alternatively, the UV/optical emission might be associated with star-formation or the narrow-line region (NLR).

\subsection{The mid-IR spectrum}
Fig.\,\ref{mirspec} shows the mid-IR spectra of IC\,3639 obtained with VLT/VISIR (in red) and \textit{Spitzer}/IRS (in light and dark green for the full-aperture and optimal-difference spectra, respectively). Three main aspects emerge from the analysis of these data: \textit{i)} the continuum emission is consistent for both instruments, in terms of flux and spectral slope, in agreement with a point-like source dominating the total mid-IR flux in the inner kpc of the galaxy; \textit{ii)} the presence of strong fine-structure emission lines corresponding to both low- (e.g. [\ion{Ne}{ii}], [\ion{S}{iv}]) and high-ionisation species (e.g. [\ion{Ne}{v}], [\ion{O}{iv}]). About $74\%$ of the [\ion{Ne}{ii}]$_{12.8\, \rm{\mu m}}$ line flux seen in the full-aperture extraction from \textit{Spitzer}/IRS is still present in the smaller slit aperture of the VISIR spectrum (see Table\,\ref{flux}), suggesting that the low-ionisation gas is mostly unresolved at a resolution of $\sim 80\, \rm{pc}$; \textit{iii)} the intense polycyclic aromatic hydrocarbon (PAH) feature at $11.3\, \rm{\micron}$ detected in the full-aperture IRS spectrum becomes a factor of $\sim 3$ fainter in both the VISIR and the optimal-difference IRS spectra (the features at $6.2$, $7.7$, $8.6$, and $17\, \rm{\micron}$ are also detected with IRS in the low-spectral resolution mode; see \citeads{2009ApJ...701..658W}, \citeads{2010ApJ...724..140D}, \citeads{2012ApJ...758....1L}). Most of the PAH$_{11.3\, \rm{\mu m}}$ emission and a fraction of the [\ion{Ne}{ii}]$_{12.8\, \rm{\mu m}}$ line flux seem to come from diffuse emission extended over an area of $4\farcs7 \times 11\farcs3$, and thus are removed in the optimal-difference IRS spectrum.

For comparison, Fig.\,\ref{mirspec} also shows the photometry derived from the VISIR NEII acquisition image, plus additional subarcsec aperture measurements in the literature (black dots, see Table\,\ref{flux}; \citeads{2009A&A...502..457G,2014MNRAS.439.1648A}). Errorbars in the horizontal and vertical axes correspond to filter \textsc{fwhm} bandwidths and flux uncertainties, respectively. For a better comparison with the imaging data, synthetic photometry was also performed on the VISIR spectrum (black squares in Fig.\,\ref{mirspec}), using the \textit{pysynphot} package included in \textsc{stsci\_python}\footnote{\textsc{stsci\_python} and \textit{PyRAF} are products of the Space Telescope Science Institute, which is operated by AURA for NASA.} and the transmission curves for: the VISIR filters\footnote{\url{http://www.eso.org/sci/facilities/paranal/instruments/visir/inst/}} PAH2 ($11.3\, \rm{\micron}$, $\Delta\lambda\,0.59\, \rm{\micron}$), NEII ($12.81\, \rm{\micron}$, $\Delta\lambda\, 0.21\, \rm{\micron}$), NEII\_2 ($13.04\, \rm{\micron}$, $\Delta\lambda\, 0.22\, \rm{\micron}$); and the T-ReCS \textit{N}-band filter ($10.36\, \rm{\micron}$, $\Delta\lambda\, 2.64\, \rm{\micron}$). Fluxes derived from imaging photometry are in agreement, within uncertainties, with the spectrophotometric measurements. Thus, possible variability in the mid-IR is not detected for the dataset analised here, which covers the period 2004--2012.

\subsection{MIDI correlated flux distribution}\label{vis}
\begin{figure}[t]
  \sidecaption
  \includegraphics[width = \columnwidth]{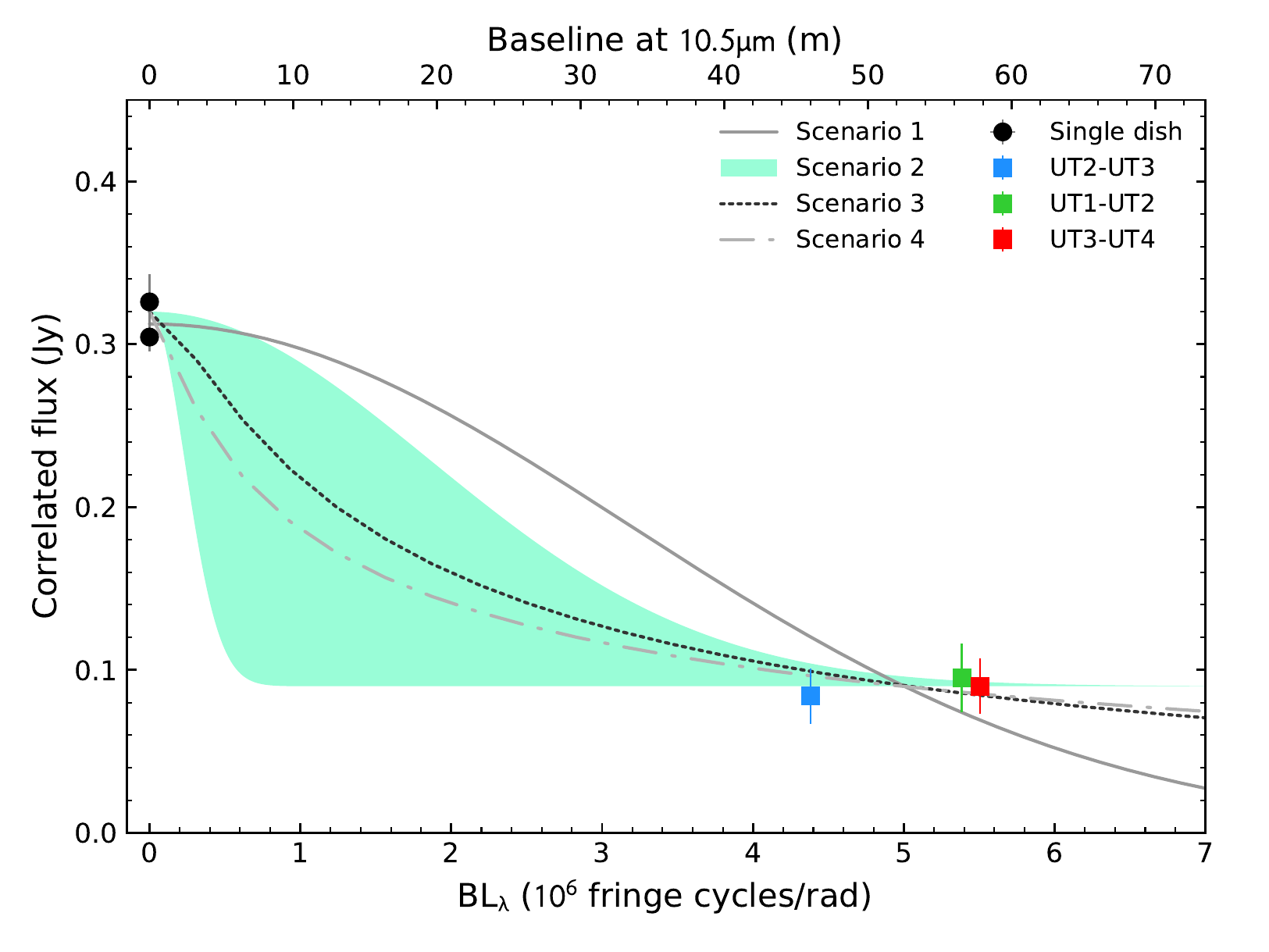}
  \caption{Black dots represent the total flux measured by single dish telescopes (VISIR/PAH2 and T-ReCS/\textit{N}-band), and filled squares represent the correlated fluxes measured by MIDI. Four possible scenarios are shown: a compact Gaussian distribution with \textsc{fwhm} $= 5.8 \pm 0.3\, \rm{pc}$ in size (grey-solid line); an unresolved source plus a partially resolved Gaussian component of $\sim 10$--$80\, \rm{pc}$ (core-halo configuration; light green-shaded area); a power-law distribution with an inner radius of $R_{\rm in} = 0.08\, \rm{pc}$, a half-light radius of $R_{1/2} = 1.8\, \rm{pc}$, an extended dust density distribution ($\alpha = 0.5$), and a temperature gradient index for ISM grains ($\beta = -0.36$; back-dotted line); a power-law distribution with an inner radius of $R_{\rm in} = 0.4\, \rm{pc}$, $R_{1/2} = 7.3\, \rm{pc}$, a flat density distribution ($\alpha = 0$), and a temperature gradient index for ISM grains ($\beta = -0.36$; grey-dot-dashed line).}\label{vis_plot}
\end{figure}

The comparison between the single-dish and the interferometric measurements provides us with a first guess of the mid-IR brightness distribution. The VISIR total flux of $\sim$\,$320\, \rm{mJy}$, coming from spatial scales of less than $0\farcs5$ (see Sect.\,\ref{data}), is much brighter than the MIDI correlated flux at short baselines ($46\, \rm{m}$, corresponding to a spatial scale of $\sim 47\, \rm{mas}$; blue square in Fig.\,\ref{mirspec}), which drops to $F_\nu = 84 \pm 17\, \rm{mJy}$. That is a factor of 3--4 times fainter than the continuum level in the VISIR spectrum, suggesting that a large fraction of the mid-IR flux is extended on a spatial scale of a few tens of parsecs. In Fig.\,\ref{vis_plot} we explore four different scenarios for the mid-IR brightness distribution to explain the MIDI correlated fluxes.

  \textbf{Scenario 1:} a single component with a Gaussian brightness distribution, which is progressively resolved with increasing baseline (grey-solid line in Fig.\,\ref{vis_plot}). In this case, the best fit of the single-dish \textit{N}-band measurements and the MIDI correlated fluxes yields a \textsc{fwhm} of $5.8 \pm 0.3\, \rm{pc}$, at the distance of IC\,3639. To estimate the fit error we used a bootstrapping Monte-Carlo routine that generates a set of synthetic data from the observed values plus a normal distribution based on the observed errors. The final fit parameters and associated errors correspond, respectively, to the average and dispersion values obtained after fitting 1000 synthetic datasets.

  \textbf{Scenario 2:} an unresolved point source plus a partially resolved Gaussian component with \textsc{fwhm} $\sim 10$--$80\, \rm{pc}$ (core-halo configuration; light green-shaded area). This scenario is motivated by the finding of such structures in several other AGN (c.f.\ \citeads{2013A&A...558A.149B}). The size of the extended component is constrained by the unresolved flux in single-dish observations ($\lesssim 80\, \rm{pc}$) and the flat visibility distribution seen by MIDI ($\gtrsim 10\, \rm{pc}$). Therefore, the unresolved component only accounts for $\sim 28\%$ of the total nuclear mid-IR flux, while $72\%$ is radiated by an extended structure of a few tens of parsecs in size, unresolved by the single-dish observations but completely resolved by MIDI (Fig.\,\ref{vis_plot}). In contrast to the compact Gaussian configuration, this scenario explains the flat distribution of the correlated fluxes seen by MIDI.

  \textbf{Scenario 3:} a single power-law distribution as described in \citetads{2007A&A...476..713K,2011A&A...536A..78K}. We assume an inner torus radius of $R_{\rm in} = 0.08\, \rm{pc}$ (from \citeads{2006ApJ...639...46S,2007A&A...476..713K}), corresponding to a nuclear UV luminosity of $L_{\rm UV} \sim 3 \times 10^{44}\, \rm{erg\,s^{-1}}$, which was derived from the $L_{\rm X} \sim 2 \times 10^{43}\, \rm{erg\,s^{-1}}$ based on \textit{NuSTAR} observations (from \citeads{2016ApJ...833..245B}, adapted to the distance used in this work) and the $L_{\rm UV}$--$L_{\rm X}$ relation from \citetads{2012A&A...539A..48M}. We further assume a sublimation temperature of $1500\, \rm{K}$, a grain size of $0.05\, \rm{\micron}$, and a dust temperature gradient index of $\beta = -0.36$, corresponding to ISM grains [$T \propto (r/R_{\rm in})^\beta$; \citeads{1987ApJ...320..537B}]. This scenario (black-dotted line in Fig.\,\ref{vis_plot}) could explain the visibilities by using a very extended dust density distribution [$\alpha = 0.5$, for $S_\nu \propto (r / R_{\rm in})^\alpha$]. This geometry implies a decreasing dust density with decreasing radius in order to reproduce the rapid drop in visibility with increasing baseline length, i.e. a torus with a lower concentration towards the centre.

  \textbf{Scenario 4:} a single power-law distribution with $R_{\rm in} = 0.4\, \rm{pc}$, which implies a $L_{\rm UV} \sim 7 \times 10^{45}\, \rm{erg\,s^{-1}}$, and a flat dust density distribution ($\alpha = 0$). In this configuration the brightness radial gradient is determined only by the temperature gradient. This model is able to reproduce the observed visibilities using a flat density distribution (grey dot-dashed line in Fig.\,\ref{vis_plot}), but requires that little or no dust emission at all should come from the innermost $\sim 0.4\, \rm{pc}$. An inner radius of roughly half a parsec would imply a very high UV luminosity ($\sim 7 \times 10^{45}\, \rm{erg\,s^{-1}}$), which is significantly larger than the upper error limit of $L_{\rm disc} < 1.3 \times 10^{45}\, \rm{erg\,s^{-1}}$ given by \citetads{2016ApJ...833..245B} at 90\% confidence. The correlated flux predicted would overestimate the MIDI measurements if either a fainter luminosity is considered (corresponding to a smaller $R_{\rm in}$), a more concentrated dust distribution is used ($\alpha < 0$), or a temperature gradient index for large grains ($\beta \sim -0.5$) is assumed.

In summary we conclude that from these four scenarios the core-halo configuration (scenario 2) seems to agree best to the observations. The power law distribution can also be accommodated to the observed visibilities, but a very large inner radius or a very extended dust density distribution should be used in order to balance the compact brightness distribution derived from the temperature gradient. As we will show in the following section, the core-halo scenario is in good agreement with the physical picture emerging when additional observational evidences are taken into account.

%~~~~~~~~~~~~~~~~~~~~~~~~~~~~~~~~~~~~~~~~~~

\begin{table*}
\caption{Mid-IR spectral lines and diagnostics measured for IC\,3639 including: line ratios, equivalent width (EW) of diverse PAH features, flux ratios of different PAH complexes, strength of the silicate absorption band at $9.7\, \rm{\micron}$ (S$_{\rm sil}$), and PAH-to-integrated continuum ratios.}\label{diag}
\centering
\begin{tabular}{lcl}
  Line/Diagnostic & Value & Ref. \\  
  \hline \\[-0.2cm]
 {[}\ion{Ne}{v}]$_{14.3}$/[\ion{Ne}{ii}]$_{12.8}$                              & $0.25 \pm 0.01$    & \citetads{2008ApJ...676..836T} \\
                                                                             & $0.33 \pm 0.02$\tablefootmark{a} & This Work \\
 {[}\ion{O}{iv}]$_{25.9}$/[\ion{Ne}{ii}]$_{12.8}$                              & $0.47 \pm 0.01$    & \citetads{2008ApJ...676..836T} \\
                                                                              & $0.63 \pm 0.03$\tablefootmark{a} & This Work \\
 {[}\ion{O}{iv}]$_{25.9}$/([\ion{Ne}{ii}]$_{12.8}$ + [\ion{Ne}{iii}]$_{15.6}$)  & $0.29 \pm 0.01$    & \citetads{2016ApJS..226...19F} \\
 {[}\ion{O}{iv}]$_{25.9}$/[\ion{O}{iii}]$_{88}$                                & $0.51 \pm 0.02$    & \citetads{2016ApJS..226...19F} \\
 EW(PAH$_{6.2\, \rm{\mu m}}$)                                                   & $0.185 \pm 0.003\, \rm{\micron}$ & \citetads{2009ApJ...701..658W} \\
 EW(PAH$_{7.7\, \rm{\mu m}}$)                                                   & $1.39\, \rm{\micron}$   & \citetads{2012ApJ...758....1L} \\
 EW(PAH$_{11.3\, \rm{\mu m}}$)                                                 & $0.39\, \rm{\micron}$   & \citetads{2012ApJ...758....1L} \\
 EW(PAH$_{17\, \rm{\mu m}}$)                                                   & $0.13\, \rm{\micron}$    & \citetads{2012ApJ...758....1L} \\
 $\frac{f_{\rm{PAH}6.2\, \rm{\mu m}}}{f_{\rm{PAH}7.7\, \rm{\mu m}}}$                & $0.14$             & \citetads{2009ApJ...701..658W}, \citetads{2012ApJ...758....1L} \\
 $\frac{f_{\rm{PAH}11.3\, \rm{\mu m}}}{f_{\rm{PAH}7.7\, \rm{\mu m}}}$              & $0.29$             & \citetads{2012ApJ...758....1L} \\
 S$_{\rm sil}$                                                                & $-0.312$            & \citetads{2009ApJ...701..658W} \\
 $\log \left( \frac{f_{PAH6.2\, \rm{\mu m}}}{f_{5.3-5.8\, \rm{\mu m}}} \right)$    & $\sim 0.007$       & This Work \\
 $\log \left( \frac{f_{14-16\, \rm{\mu m}}}{f_{5.3-5.8\, \rm{\mu m}}} \right)$     & $\sim 1.05$        & This Work \\[0.3cm]
\hline                  
\end{tabular}
\tablefoot{
\tablefoottext{a}{Using the [\ion{Ne}{ii}]$_{12.8\, \rm{\mu m}}$ flux measured in the VISIR spectrum.}
}
\end{table*}

\section{Discussion}\label{discussion}

The main questions that arise on the nature of the nuclear emission in IC\,3639 are: \textit{i)} the relative contribution of the AGN and the starburst to the mid-IR emission; and \textit{ii)} the physical properties of these two components, i.e. their size, geometry, and spectral shape. Our approach is to infer the relative AGN and starburst contributions using IR diagnostics based on fine-structure emission lines, the continuum shape, and spectral features such as PAH and the silicate band at $9.7\, \rm{\micron}$ (Section\,\ref{iso}), and compare these results with the analysis of the correlated flux distribution and the size of the mid-IR components resolved by MIDI (Section\,\ref{agn_stb}). Finally, we also use the multiwavelength spectral energy distribution (SED) to shed light on the nature of the nuclear emission (Section\,\ref{sed}).

\subsection{IR spectral features and colours}\label{iso}
The relative contributions from the AGN and the starburst to the gas photoionisation can be inferred from the IR fine-structure emission lines. Both the AGN and the starburst are able to produce strong lines of low-ionisation species --\,e.g. [\ion{Ne}{ii}]$_{12.8\, \rm{\mu m}}$ and [\ion{S}{iv}]$_{10.5\, \rm{\mu m}}$\,-- however the AGN is the dominant or the only contribution to emission lines of high-ionisation species, like [\ion{Ne}{v}]$_{14.3\, \rm{\mu m}}$ and [\ion{O}{iv}]$_{25.9\, \rm{\mu m}}$. Thus, line ratios of high- to low-ionisation species provide a reliable estimate of the relative contribution of these two components \citepads{2002A&A...393..821S,2008ApJ...689...95M}. From the mid-IR line fluxes measured by \citetads{2008ApJ...676..836T} in the \textit{Spitzer}/IRS-SH and LH spectra listed in Table\,\ref{flux}, we obtain the line ratios of [\ion{O}{iv}]$_{25.9}$/[\ion{Ne}{ii}]$_{12.8}$\,$ = 0.47 \pm 0.01$ and [\ion{Ne}{v}]$_{14.3}$/[\ion{Ne}{ii}]$_{12.8}$\,$ = 0.25 \pm 0.01$ (see Table\,\ref{diag}). These values are far from those obtained in AGN-dominated nuclei (with typical ratios of [\ion{O}{iv}]$_{25.9}$/[\ion{Ne}{ii}]$_{12.8}$\,$\gtrsim 1.4$ and [\ion{Ne}{v}]$_{14.3}$/[\ion{Ne}{ii}]$_{12.8}$ $\gtrsim 0.6$; \citeads{2002A&A...393..821S}), but similar to the ratios found in ULIRGs, usually associated with strong star formation \citepads{2006ApJ...640..204A,2007ApJ...656..148A}. Following \citetads{2002A&A...393..821S}, a simple linear mixing model suggests an AGN contribution of $\sim$\,$17\%$ from the [\ion{O}{iv}]$_{25.9}$/[\ion{Ne}{ii}]$_{12.8}$ ratio and $\sim$\,$23\%$ from the [\ion{Ne}{v}]$_{14.3}$/[\ion{Ne}{ii}]$_{12.8}$ ratio, in agreement with the results obtained by \citetads{2008ApJ...689...95M} for the case of IC\,3639. This result is supported by photoionisation models for AGN with ionisation parameter values in the $-2.5 < \log U < -1.5$ range, which predict a $\sim$\,$30\%$ AGN contribution for [\ion{Ne}{v}]$_{14.3}$/[\ion{Ne}{ii}]$_{12.8} \sim 0.3$ \citepads{2008ApJ...678..686A}. Additional line ratios such as [\ion{O}{iv}]$_{25.9}$/([\ion{Ne}{ii}]$_{12.8}$ + [\ion{Ne}{iii}]$_{15.6}$) and [\ion{O}{iv}]$_{25.9}$/[\ion{O}{iii}]$_{88}$ (listed in Table\,\ref{diag}), which are more robust diagnostics for the relative AGN/starburst contribution under harder radiation fields (e.g. \citeads{1998A&A...333L..75L}, \citeads{2016ApJS..226...19F}), also suggest a dominant contribution of the starburst component to the low-ionisation emission lines.

The line ratio diagnostics still apply if we use the [\ion{Ne}{ii}]$_{12.8\, \rm{\mu m}}$ line flux measured in the VISIR spectrum, which contains about $74\%$ of the flux seen by \textit{Spitzer}/IRS for the same emission line in a much wider aperture (see Table\,\ref{flux}). In this case, the relative AGN contribution would increase to $\sim$\,$23\%$ from the [\ion{O}{iv}]$_{25.9}$/[\ion{Ne}{ii}]$_{12.8}$ ratio and $\sim$\,$30\%$ from the [\ion{Ne}{v}]$_{14.3}$/[\ion{Ne}{ii}]$_{12.8}$ ratio, assuming the same linear mixing model as before. We note that the IR fine-structure lines have not been corrected by extinction, however the detection of [\ion{Ne}{v}]$_{3426\, \rm{\AA}}$ emission in the optical spectrum (see fig.\,9b in \citeads{1995ApJS...98..103S}) suggests that the NLR emission in this galaxy is not strongly affected by dust, thus the effect of extinction in the mid- to far-IR range is expected to be negligible.

In order to further test the relative contribution of the starburst component, we used additional diagnostics based on IR spectral features (see Table\,\ref{diag}) and the shape of the continuum emission measured for IC\,3639. The mid-IR colour $W1 - W2 = 0.87$ ($3.4$ to $4.6\, \rm{\micron}$), taken from the Wide-field Infrared Survey Explorer (\textit{WISE}) point-source catalogue, falls within the limit defined by \citetads{2012ApJ...753...30S} for AGN-heated dust ($W1 - W2 > 0.8$, $W2 < 15$). However, the $W2 - W3 = 4.27$ ($4.6$ to $11.6\, \rm{\micron}$) and $W3 - W4 = 3.10$ ($11.6$ to $22.1\, \rm{\micron}$) values place IC\,3639 outside of the colour wedge defined by \citetads{2012MNRAS.426.3271M} for luminous AGN, and rather close to the mid-IR colours of M82 or Arp\,220. A more detailed analysis of the continuum emission for IC\,3639 has been performed recently by \citetads{2016MNRAS.458.4297G}, using a broad-band UV to far-IR SED decomposition technique. The latter combines spectral templates of the dusty torus, the accretion disc, and various stellar populations (see \citeads{2013A&A...551A.100B} for further details). These authors find an AGN fraction of $\sim$\,$0.3 \pm 0.6$ for the nucleus in the $5$--$40\, \rm{\micron}$ range, in agreement with our results albeit with very large uncertainty.

Finally, we analyse other IR spectral features such as the PAH and silicate bands. In this regard, the equivalent width of the PAH complex at $6.2\, \rm{\micron}$ is consistent with a mixed contribution of both AGN and star formation \citepads{2011ApJ...730...28P}. A further analysis comparing the ratios of the PAH$_{6.2\, \rm{\mu m}}$ flux to the $5.3$--$5.8\, \rm{\micron}$ continuum and the continuum ratio between $14$--$16\, \rm{\micron}$ and $5.3$--$5.8\, \rm{\micron}$ (following \citeads{2000A&A...359..887L}, \citeads{2007ApJ...656..148A}, and \citeads{2009ApJS..182..628V}; values listed in Table\,\ref{diag}), suggests that the heating of both PAH and dust grains is consistent with a combination of $\sim$\,$50\%$ AGN and a mixture of emission from \textsc{H\,ii} and photo-dissociation regions (PDR) for the remaining half fraction. This is also in agreement with the relatively high degree of PAH excitation inferred from the PAH ratios $F_{6.2\, \rm{\mu m}}$/$F_{7.7\, \rm{\mu m}}$ vs. $F_{11.3\, \rm{\mu m}}$/$F_{7.7\, \rm{\mu m}}$ \citepads{2014ApJ...790..124S}. We note that the diagnostics based on PAH features, specially those using the IRS-SL low-resolution spectra (slit width of $3\farcs6 \times 57''$), are probably dominated by the extended emission with little contribution from the nuclear unresolved source, as shown by the comparison between the full-aperture and the optimal-difference IRS-HR spectra (light green and dark green lines in Fig.\,\ref{mirspec}, respectively). Finally, the EW(PAH$_{6.2\, \rm{\mu m}}$) was compared to the strength of the $9.7\, \rm{\micron}$ silicate feature, S$_{\rm Sil}$, following the diagnostics in \citetads{2007ApJ...654L..49S}. The latter places IC\,3639 in the intermediate region between galaxies with a mid-IR spectrum dominated by AGN-heated dust and PAH-dominated spectra typical of starburst galaxies, with EW(PAH$_{6.2\, \rm{\mu m}}$) and S$_{\rm Sil}$ similar to other Seyfert nuclei in LIRGs with a strong contribution from star formation, such as NGC\,7469 \citepads{1995ApJ...444..129G}.

\subsection{A compact AGN/starburst nucleus?}\label{agn_stb}
For a mid-IR luminosity of a few times $10^{42}\, \rm{erg\,s^{-1}}$, the expected size distribution of the torus is about $1\, \rm{pc}$, based on interferometric observations of nearby AGN \citepads{2009A&A...502...67T,2011A&A...531A..99T,2013A&A...558A.149B}. This is in agreement with the core-halo configuration shown in Fig.\,\ref{vis_plot}, in which the AGN dominates the unresolved emission seen by MIDI. Extended nuclear emission in the mid-IR has been detected for other nearby AGN, for instance, an extended component with a size of $\sim$\,$200\, \rm{pc}$ and a temperature of $\sim$\,$165\, \rm{K}$ was reported for the nucleus of NGC\,4151 by \citetads{2003ApJ...587..117R}. However, these authors conclude that the extended emission can be ascribed to dust grains in the NLR and not to the torus. This is in line with the detection of mid-IR emission extended along the axis direction of several AGN over scales of few tens to few hundreds of parsecs in size (also known as polar dust; \citeads{2013ApJ...771...87H}, \citeads{2016ApJ...822..109A}), but also at subparsec to few parsecs scales in mid-IR interferometric observations (e.g. \citeads{2013ApJ...771...87H}, \citeads{2014A&A...563A..82T}, \citeads{2014A&A...565A..71L}). A framework to explain these observations has been proposed recently by \citetads{2017ApJ...838L..20H}, where the torus is modelled as a compact disc plus an dusty wind blown by the active nucleus along the axis of the system. In the case of IC\,3639, the extended dusty wind could possibly explain the extended component resolved by MIDI in the mid-IR. However, in order to reconcile this scenario with the measured line ratios, a considerable contribution from star formation should occur somewhere within the disc-plus-wind structure. Alternatively, an uncommon weak ionisation radiation field from the AGN with an ionisation parameter of $\log U \lesssim -3$ would have to be invoked in order to reproduce a [\ion{Ne}{v}]$_{14.3}$/[\ion{Ne}{ii}]$_{12.8}$ ratio of $\sim$\,$0.3$ without additional contribution from star formation activity to the photoionised gas \citepads{2008ApJ...678..686A}. Finally, the three correlated fluxes measured by MIDI along different baseline orientations (see right panel in Fig.\,\ref{mirmap}) are consistent, thus there is no evidence for a possible extension along a preferential direction.

The radial mid-IR brightness distribution seen by MIDI is flat between $46$ and $58\, \rm{m}$. Thus, a single Gaussian component does not provide a good fit to the visibilities, as shown in Fig.\,\ref{vis_plot}. A single power-law distribution can reproduce the estimated visibilities, but only after imposing restrictions on the geometry: either the dust density profile decreases steeply towards the AGN with decreasing radius [$\propto (r/R_{\rm in})^{0.5}$; dotted-black line in Fig.\,\ref{vis_plot}], in contrast with other AGN observed with mid-IR interferometry ($\alpha \lesssim 0$, \citeads{2011A&A...536A..78K}); or the inner radius is significantly larger as expected from the AGN luminosity for this galaxy ($\gtrsim 0.4\, \rm{pc}$; dot-dashed grey line). Both scenarios imply a lower concentration of dust in the innermost region around the AGN to balance for the steep brightness profile imposed by the temperature gradient ($\beta = -0.36$ for ISM dust). If the temperature distribution is steeper when compared to ISM grains (e.g. $\beta = -0.5$ for large grains), an even less concentrated dust distribution (larger $\alpha$ and/or $R_{\rm in}$) would be required to fit the visibilities measured by MIDI. Therefore, none of these alternatives seem to be an appealing scenario to describe the mid-IR brightness distribution in IC\,3639.

We favour the core-halo configuration (scenario 2), represented in Fig.\,\ref{vis_plot} by an extended Gaussian ($10$--$80\, \rm{pc}$) plus an unresolved source (light green-shaded area). The extended component dominates the total mid-IR emission ($72\%$) and would be associated to dust heated by star formation, while the unresolved component ($28\%$ of the total mid-IR flux) is associated to the AGN. This result is in good agreement with diagnostics based on IR fine-structure lines: the strength of low-excitation lines, such as [\ion{Ne}{ii}]$_{12.8\, \rm{\mu m}}$, compared to high-excitation transitions ([\ion{Ne}{v}]$_{14.3\, \rm{\mu m}}$, [\ion{O}{iv}]$_{25.9\, \rm{\mu m}}$) suggests that star formation contributes about $67$--$70\%$ to the gas ionisation, with a remaining $23$--$30\%$ due to AGN activity (Section\,\ref{iso}). Furthermore, similar values for the relative AGN/starburst contributions are derived from the decomposition of the mid-IR continuum, as shown by \citetads{2016MNRAS.458.4297G}. For these reasons we favour the extended core-halo configuration shown in Fig.\,\ref{vis_plot} as the most consistent explanation for the characteristics of the brightness distribution, the ionised gas, and the shape of the IR continuum emission in IC\,3639. In this scenario, the unresolved component ($\sim$\,$30\%$ of the flux) corresponds to the AGN heated dust in the innermost few parsecs ($\lesssim 10\, \rm{pc}$), while most of the IR emission (70\%) is associated to dust heated by the starburst component, extended over a spatial scale of $\sim$\,$10$--$80\, \rm{pc}$.

\begin{figure*}[t]
  \sidecaption
  \includegraphics[width = 12cm]{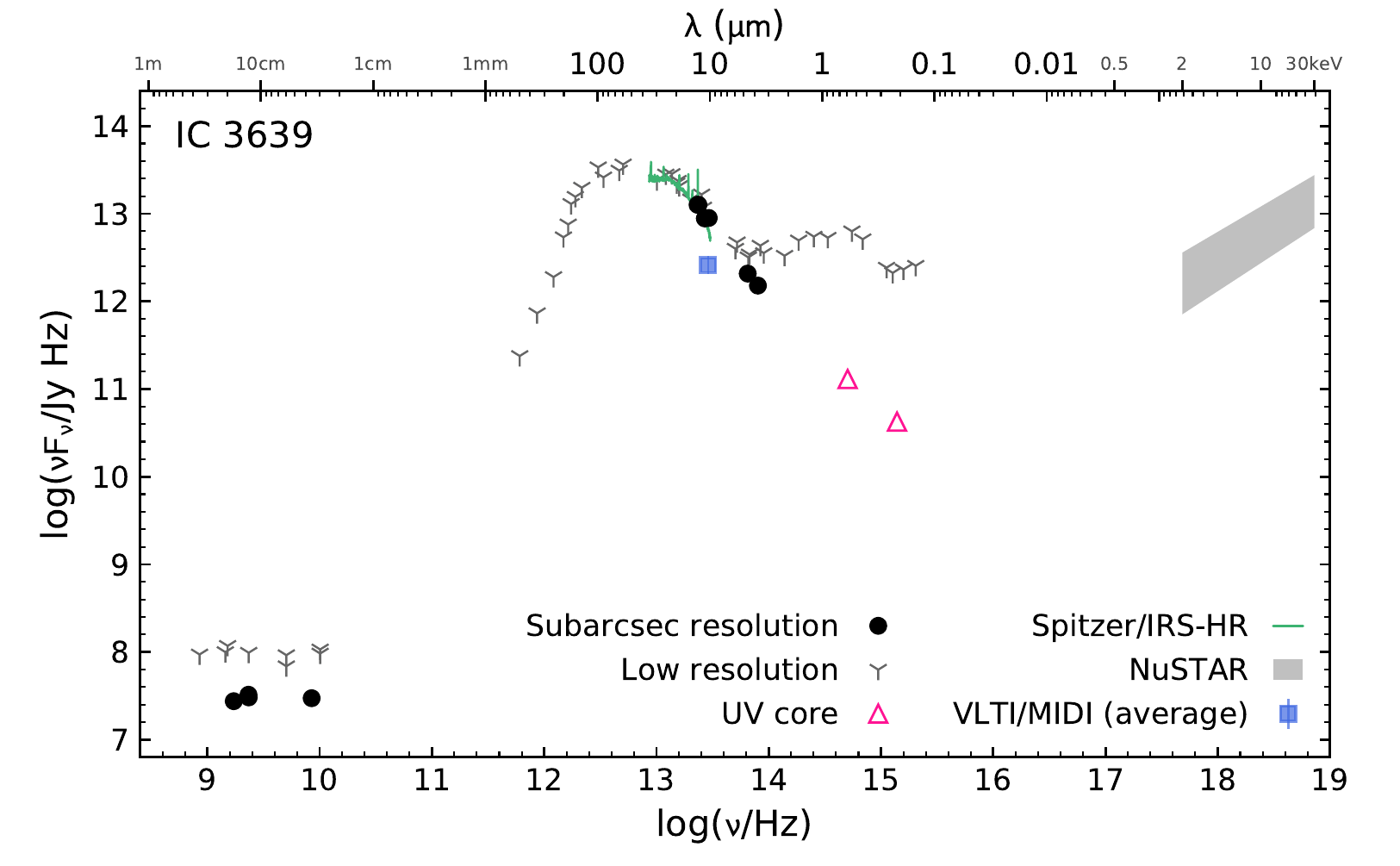}
  \caption{SED for the nucleus of IC\,3639 at a $\lesssim 0\farcs4$ resolution (black dots), compared with the MIDI measurement (blue square) and the \textit{Spitzer}/IRS-HR optimal-difference spectrum from the CASSIS atlas (dark green line). Grey spikes represent the low-spatial resolution measurements, i.e. the integrated SED of the host galaxy. Violet triangles correspond to the UV core. The grey-shaded area indicates the $2$--$10\, \rm{keV}$ and $0.5$--$30\, \rm{keV}$ intrinsic fluxes estimated for the models T and M in \citetads{2016ApJ...833..245B}.\vspace{2.4cm}}\label{sed_plot}
\end{figure*}

However, the case of IC\,3639 is remarkable due to the compactness of the whole system: a strong starburst and an active nucleus coexist in the central $80\, \rm{pc}$ of the galaxy. This scenario is consistent with a nuclear SFR of $\sim$\,$0.26\, \rm{M_\odot\,yr^{-1}}$ derived from the VISIR PAH$_{11.3\, \mu m}$ flux in Table\,\ref{flux} and the linear fit calibration given by \citetads{2016ApJ...818...60S}, although it is in contrast with the much lower value of $\sim$\,$5 \times 10^{-3}\, \rm{M_\odot\,yr^{-1}}$ estimated from the UV luminosity in the filter F210M (from Table\,\ref{flux} and the NUV calibration in \citeads{2012ARA&A..50..531K}). Thus, a large fraction of the SFR might be obscured in the optical/UV range. \citetads{2017ApJ...835...91L} suggest that a strong starburst contribution might have an important role to explain the large equivalent width of the Fe K$\alpha$ X-ray line. Therefore the starburst component could be responsible for a significant part of the obscuration of the AGN, supporting also the geometrical thickness of the torus structure through stellar winds and SNe explosions (\mbox{\citeads{2002ApJ...566L..21W}}, \mbox{\citeads{2008A&A...491..441V}}, \mbox{\citeads{2009MNRAS.393..759S}}, \mbox{\citeads{2012ApJ...752...87A}}). The star formation activity could be feeding a moderate BH ($7.3 \times 10^6\, \rm{M_\odot}$; \citeads{2007ApJ...660.1072W}), growing at a rate of $\log(L_{\rm bol}/L_{\rm edd}) \sim -1$ ($L_{\rm bol} \sim 8\, \times 10^{43}\, \rm{erg\,s^{-1}}$ derived from the flux measured with MIDI at $\sim$\,$10\, \rm{\micron}$ and the $L_{12\, \rm{\micron}}$--$L_{\rm bol}$ relation from \citeads{1995ApJ...453..616S} for Seyfert galaxies in the 12 micron sample). IC\,3639 is one of the most compact nuclear starbursts embedded in dust and surrounding an AGN, comparable to the case of NGC\,4418. The latter has a brighter IR core ($L_{\rm bol} \sim 4 \times 10^{44}\, \rm{erg\,s^{-1}}$) comparable in size ($\lesssim 20\, \rm{pc}$; \citeads{2013ApJ...764...42S}, \citeads{2013A&A...556A..66C}), although affected by a much heavier obscuration in the mid-IR range ($S_{\rm sil} \sim -4$; \citeads{2001A&A...365L.353S}, \citeads{2007ApJ...654L..49S}).

\subsection{Spectral energy distribution}\label{sed}

The analysis of the SED gives us information on the mechanisms that produce the observed emission in the nucleus of IC\,3639. A multiwavelength SED with subarcsec resolution (black dots in Fig.\,\ref{sed_plot}) is built using the flux measurements given in Table\,\ref{flux} plus comparable high-spatial resolution data in radio and X-ray wavelengths compiled from the literature \citepads{1995MNRAS.276.1373S,1994ApJ...432..496R,2000MNRAS.314..573T,2005A&A...444..119G}. A low-spatial resolution SED based on measurements with larger apertures --several arcsecs to few arcmin-- was also compiled (grey spikes), including measurements from NED\footnote{The NASA/IPAC Extragalactic Database (NED) is operated by the Jet Propulsion Laboratory, California Institute of Technology, under contract with the National Aeronautics and Space Administration.} as well as the \textit{Herschel}/SPIRE \citepads{2013ApJ...768...55P}, \textit{Akari} \citepads{2011PASP..123..852Y} and \textit{WISE} point source catalogues. In all cases, UV, optical, and IR fluxes are corrected from Galactic extinction for a reddening of $E(B - V) = 0.06\, \rm{mag}$ \mbox{\citepads{2011ApJ...737..103S}}, using the extinction law from \citetads{1989ApJ...345..245C}. For clarity, broad-band measurements including PAH features are not included in the low-spatial resolution SED. When compared with subarcsec resolution measurements, the low-spatial resolution SED reveals those spectral ranges dominated by the host galaxy (e.g. optical, radio) from those dominated by the nuclear/circumnuclear emission (mid-IR).

In this context, the interferometric data provide a key point to interpret the origin of the nuclear emission in IC\,3639. About 70\% of the mid-IR flux detected by VISIR originates in a scale of $\gtrsim 10$--$80\, \rm{pc}$, and its continuum emission traced by the \textit{Spitzer}/IRS spectrum peaks at $20$--$40\, \rm{\micron}$ (see Fig.\,\ref{sed_plot}). For the case of thermal emission, the latter would correspond to a temperature value in the $70$--$150\, \rm{K}$ range. ISAAC measurements suggest the presence of hotter dust ($\sim$\,$320\, \rm{K}$) in the nuclear region. This range could be dominated by the AGN emission, but future \textit{L}- and \textit{M}-band interferometric observations with MATISSE (Multi AperTure mid-Infrared SpectroScopic Experiment) would be needed in order to probe this scenario. The shape of the SED shows a steep decline with increasing frequency (see Fig.\,\ref{sed_plot}), but it is not clear whether the UV/optical continuum is originated in the AGN itself or by a young starburst plus an underlying old stellar population as suggested by \citetads{1998ApJ...505..174G}. In this case the UV luminosity of the core represents only $\sim$\,$1\%$ of the total luminosity reprocessed in the mid-IR range. This is in agreement with the Balmer decrement measured in IC\,3639 (H$\alpha$/H$\beta$ = 6.1), which corresponds to an extinction value of $A_{\rm V} \sim 2.2\, \rm{mag}$, enough to reconcile the absorption-corrected UV luminosity ($L_{\rm UV} \sim 8 \times 10^{43}\, \rm{erg\,s^{-1}}$) with the total mid-IR luminosity.

Alternatively, the nuclear UV emission and the extended lobes in Fig.\,\ref{mirmap} might be produced by the NLR and/or scattered light from the nucleus. This would be in agreement with the extended emission in the soft X-rays seen along the East--West direction by \citetads{2007ApJ...656..105G}. In order to avoid confusion, we use a different symbol for the nuclear UV/optical flux measurements in Fig.\,\ref{sed_plot} (violet triangles).

%%%%%%%%%%%%%%%%%%%%%%%%%%%%%%%%%%%%%%%%%%%%%%%%%%%%%%%%%%%%%%

\section{Summary}\label{summary}
In this work we combined interferometric observations at milliarcsec resolution in the mid-IR with a multi-wavelength dataset at subarcsec resolution. This approach allowed us to probe the physical scales associated with the nuclear emission in IC\,3639. This Seyfert 2 galaxy shows a strong point-like source at the resolution of VISIR ($\sim$\,$320\, \rm{mJy}$ within $\lesssim 340\, \rm{mas}$), but surprisingly most of the emission is spatially resolved in the MIDI data, showing correlated fluxes of only $\sim$\,$90\, \rm{mJy}$, that is a factor $3$ to $4$ times lower when compared to the single dish measurements.

In order to explain this large drop in visibilities, we use a core-halo model with two components: an extended one with a few tens of parsecs in size ($40$--$340\, \rm{mas}$, $\sim$\,$10$--$80\, \rm{pc}$), and an unresolved core which accounts for the correlated fluxes measured by MIDI, i.e. $\sim$\,$30\%$ of the total flux. A single dust component with a power law distribution could reproduce the observed visibilities, but a very extended dust distribution ($\alpha \gtrsim 0.4$ and/or $R_{\rm in} \gtrsim 0.4\, \rm{pc}$), uncommon among other AGN observed with mid-IR interferometry, should be assumed. Alternatively, the presence of polar dust extended along the axis of the system could possibly explain the mid-IR brightness distribution in IC\,3639.

From our study we conclude that the extended nuclear mid-IR component resolved by MIDI ($\sim$\,$10$--$80\, \rm{pc}$) is not likely associated with the AGN for the following reasons: \textit{i)} its size is too large for a torus as discussed in Sect.\,\ref{vis}; \textit{ii)} several line ratios such as [\ion{Ne}{v}]$_{14.3}$/[\ion{Ne}{ii}]$_{12.8}$, [\ion{O}{iv}]$_{25.9}$/[\ion{Ne}{ii}]$_{12.8}$, [\ion{O}{iv}]$_{25.9}$/[\ion{O}{iii}]$_{88}$, and [\ion{O}{iv}]$_{25.9}$/([\ion{Ne}{ii}]$_{12.8}$ + [\ion{Ne}{iii}]$_{15.6}$) suggest an AGN-starburst composite nucleus dominated by the starburst component; \textit{iii)} the relative AGN-starburst contribution of $\sim$\,$30$--$70\%$, inferred from the mid-IR brightness distribution in the core-halo scenario, is in very good agreement with the relative contribution of the AGN-starburst components inferred from the ionisation of the IR fine-structure lines, and the shape of the IR continuum (Section\,\ref{iso}).

Therefore, we favour the core-halo scenario as the most consistent explanation to describe the observed properties of the mid-IR nucleus in IC\,3639. The total mid-IR emission is dominated by an extended component ascribed to star formation within the innermost $80\, \rm{pc}$. The presence of an AGN is needed in order to explain the hard X-ray emission, the high-ionisation fine-structure lines in the spectrum, and the unresolved component detected by the MIDI interferometric observations. Furthermore, the proposed scenario for this galaxy could in fact be relatively common in other Seyfert galaxies. A large number of type 2 nuclei show strong point-like emission in the mid-IR at subarcsec scales, and this is commonly assumed to be the emission from the torus. Only the high-spatial resolution achieved with VLTI/MIDI, combined with the multi-wavelength analysis at subarcsec scales, allowed us to disentangle the emission components in nucleus of IC\,3639. Our results also highlight the difficulty when trying to separate the emission components or determine torus covering fractions from larger aperture SEDs as obtained with space telescopes. The next generation of interferometric instruments such as MATISSE at VLTI, would be of great help to disentangle the different contributions to the mid-IR flux in the innermost parsecs of nearby AGN.

\begin{acknowledgements}
This research made use of Astropy, a community-developed core Python package for Astronomy (Astropy Collaboration, 2013). This research has made use of the NASA/IPAC Infrared Science Archive, which is operated by the Jet Propulsion Laboratory, California Institute of Technology, under contract with the National Aeronautics and Space Administration.

J.A.F.O. acknowledges financial support from the Spanish Ministry of Economy and Competitiveness (MINECO) under grant number MEC-AYA2015-53753-P.
\end{acknowledgements}

%%%%%%%%%%%%%%%%%%%%%%%%%%%%%%%%%%%%%%%%%%%%%%%%%%%%%%%%%%%%%%
\bibliography{ic3639}
\bibliographystyle{aa}
%%%%%%%%%%%%%%%%%%%%%%%%%%%%%%%%%%%%%%%%%%%%%%%%%%%%%%%%%%%%%%
\end{document}